\documentclass{article}

\usepackage{arxiv}

\usepackage[utf8]{inputenc} 
\usepackage[T1]{fontenc}    
\usepackage{hyperref}       
\usepackage{url}            
\usepackage{booktabs}       
\usepackage{amsfonts}       
\usepackage{nicefrac}       
\usepackage{microtype}      
\usepackage{lipsum}		
\usepackage{graphicx}
\usepackage{amsmath}

\title{Dependent Conditional Value-at-Risk \\for Aggregate Risk Models}

\author{ \href{https://orcid.org/0000-0000-0000-0000}{\includegraphics[scale=0.06]{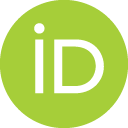}\hspace{1mm}Bony ~Josaphat}\\
	Statistics Research Division\\
	Institut Teknologi Bandung\\
	Jl. Ganesha No.10, Bandung 40132\\
	\texttt{b\_marbun@s.itb.ac.id} \\
	\And
	\href{https://orcid.org/0000-0000-0000-0000}{\includegraphics[scale=0.06]{orcid.png}\hspace{1mm}Khreshna ~Syuhada}\\
	Statistics Research Division\\
	Institut Teknologi Bandung\\
	Jl. Ganesha No.10, Bandung 40132\\
	\texttt{khreshna@math.itb.ac.id} \\
}

\hypersetup{
pdftitle={Dependent Conditional Value-at-Risk for Aggregate Risk Models},
pdfsubject={q-bio.NC, q-bio.QM},
pdfauthor={David S.~Hippocampus, Elias D.~Striatum},
pdfkeywords={First keyword, Second keyword, More},
}

\begin{document}
\maketitle

\begin{abstract}
Risk measure forecast and model have been developed in order to not only provide better forecast but also preserve its (empirical) property especially coherent property. Whilst the widely used risk measure of Value-at-Risk (VaR) has shown its performance and benefit in many applications, it is in fact not a coherent risk measure. Conditional VaR (CoVaR), defined as mean of losses beyond VaR, is one of alternative risk measures that satisfies coherent property. There has been several extensions of CoVaR such as Modified CoVaR (MCoVaR) and Copula CoVaR (CCoVaR). In this paper, we propose another risk measure, called Dependent CoVaR (DCoVaR), for a target loss that depends on another random loss, including model parameter treated as random loss. It is found that our DCoVaR outperforms than both MCoVaR and CCoVaR. Numerical simulation is carried out to illustrate the proposed DCoVaR. In addition, we do an empirical study of financial returns data to compute the DCoVaR forecast for heteroscedastic process.
\end{abstract}


\section{Introduction}
Risk measure forecast has been one of major interests in finance and insurance and developed by academia and practitioners. The common and widely used risk measure is Value-at-Risk (VaR), see e.g. McNeil et al. (2005), Kabaila and Mainzer (2018), Syuhada et al. (2020); Nieto and Ruiz (2016) provided latest review on VaR and its backtesting. It forecasts maximum tolerated risk at certain level of significance. Basically, VaR is calculated through the quantile of its loss distribution. Whilst the widely used risk measure of VaR has shown its performance and benefit in many applications, it is in fact not a coherent risk measure.

There have been some efforts done by authors to seek an improvement of VaR, beside describing formulas of VaR and CoVaR as shown in Nadarajah et al. (2016). Their works were derived in two different directions. The first is improvement of VaR forecast accuracy i.e. the coverage probability of VaR forecast is closer to the target nominal or significant level. The example of this is an improved VaR in which the method was developed by Kabaila and Syuhada (2008, 2010) and Syuhada (2020) whilst estimating confidence region by adjusted empirical likelihood to obtain better coverage was proposed by Yan and Zhang (2016). Furthermore, Kabaila and Mainzer (2018) considered linear regression model that consists of approximate VaR and exact VaR in which the former is an unbiased estimator for the latter.

The second improvement to VaR is seeking alternative risk measure(s) that capture coherent property. The commonly used coherent risk measure is the Conditional VaR (CoVaR), defined as mean of losses beyond VaR, see e.g. Artzner et al. (1999), McNeil et al. (2005), Jadhav et al. (2009, 2013), Righi and Ceretta (2015), and Brahim et al. (2018). Several extensions of CoVaR have proposed. Jadhav et al. (2013) has modified CoVaR by introducing fixed boundary, instead of infinity, for values beyond VaR. They named the risk measure as Modified CoVaR (MCoVaR). Meanwhile, another extension of CoVaR, called Copula CoVaR (CCoVaR), was suggested by Brahim et al. (2018) in which they forecast a target risk by involving another dependent risk or associate risk. The use of Copula in this dependent case is crucial. The application of this method may be found when we forecast risk premia (as a target risk) that depends on claim size (as an associate risk). Note that Kang et al. (2019) considered such premium and claim size dependence to forecast VaR and CoVaR by involving Copula.

Motivated by the work of Jadhav et al. (2013) and Brahim et al. (2018), in this paper, we propose an alternative coherent risk measure that is not only ``considering a fixed upper bound of losses beyond VaR'' but also ``taking into account a dependent risk''. Our proposed risk measure is called Dependent CoVaR (DCoVaR). When we compute an MCoVaR forecast, it will reduce number of losses beyond VaR and thus make this forecast smaller than the corresponding CoVaR. We argue that this forecast must also be accompanied by a dependent risk since this risk scenario occurs in practice, see for instance Zhang et al. (2018) and Kang et al. (2019).

This paper is organized as follows. Section 2 describes our proposed risk measure of DCoVaR in which its formula relies on joint distribution either classical or Copula. Properties of DCoVaR are also stated. The DCoVaR forecast for Pareto random loss is explained in Section 3. Such forecast is computed for target risk of Pareto and associate risk of Pareto as well. Farlie-Gumbel-Morgenstern and Archimedean Copulas are employed. The target risk may be extended to an aggregate risk. Numerical simulation is carried out. Section 4 considers a real application of DCoVaR forecast for financial returns data (NASDAQ and TWIEX) in which such returns are modeled by heteroscedastic process of GARCH. Appendix contains all proofs.

\section{Description of Dependent CoVaR forecast}

Suppose that an aggregate loss model $S_{N-k}$ is constructed by a collection of dependent random losses $X_1, X_2, \ldots, X_{N-k}$ given by $S_{N-k}=X_1+\ldots+X_{N-k}$, for $k=0,1,2\ldots,N-1$. The VaR forecast of $S_{N-k}$, at a significant level $\alpha$, is obtained by the inverse of distribution function of $S_{N-k}$, i.e. ${\rm VaR}_\alpha(S_{N-k})=F^{-1}_{S_{N-k}}(\alpha)=Q_{\alpha}$. In practice, the parameter of the model must be estimated from data. Thus, the coverage probability of this VaR forecast is bounded to $O(n^{-1})$ since it takes into account the parameter estimation error. Provided VaR forecast, $Q_{\alpha}$, the mean of losses beyond VaR to infinity may be calculated, called Conditional VaR (CoVaR). Unlike VaR, the CoVaR forecast preserves subadditivity (thus satisfies coherent property) that makes diversification reasonable. Furthermore, as stated by Koji and Kijima (2003), any coherent risk measure can be represented as a convex combination of CoVaR.

We aim to find a risk measure forecast that calculates the mean of $S_{N-k}$ beyond its VaR up to a fixed value of losses and the $S_{N-k}$ depends on another dependent or associate random loss. Our proposed risk measure forecast, namely Dependent Conditional VaR (DCoVaR), calculates the mean of $S_1$ in which $Q_{\alpha} \le S_{N-k} \le Q_{\alpha_1}$ and $S_{N-k}$ depends on another random loss $Y$ as follows
\begin{equation}\label{DCoVaR}
{\rm DCoVaR}\Big( S_{N-k} \, \Big| \, Y \Big) = E\Big[ S_{N-k} \, \Big| \, Q_{\alpha} \le S_{N-k} \le Q_{\alpha_1}, Q_{\delta}(Y) \le Y \le Q_{\delta_1}(Y) \Big], \end{equation}
where $\alpha_1=\alpha+(1-\alpha)^{a+1}$ and $\delta_1=\delta+(1-\delta)^{d+1}$ for a specified $a$ and $d$. Note that such random loss $Y$ may be (i) a single component of $S_{N-k}$, (ii) another aggregate risk model $S_{N-l}$, or (iii) a parameter model. Note also that in many applications, the distribution of $S_{N-k}$ and $Y$ may be either non-normal or not specified so that we need a Copula. In what follows, we state our proposed DCoVaR in the two propositions below.

\bigskip \noindent \textsc{Proposition 1} \\
Let $S_{N-k}$ and $Y$ be two random losses with a joint probability function $f_{S_1,Y}$. Let $\alpha,\delta \in (0,1)$. The Dependent Conditional VaR (DCoVaR) of $S_{N-k}$ given values beyond its VaR up to a fixed value of losses and a random loss $Y$ is given by
\begin{equation}
{\rm DCoVaR}_{(\alpha,a)}^{(\delta,d)} \Big( S_{N-k} \, \Big| \, Y \Big) = \frac{\int\limits_{Q_{\delta}}^{Q_{\delta_1}} \, \int\limits_{Q_{\alpha}}^{Q_{\alpha_1}} \, s \, f_{S_{N-k},Y} (s,y) \, ds \, dy}{\int\limits_{Q_{\delta}}^{Q_{\delta_1}} \, \int\limits_{Q_{\alpha^1}}^{Q_{\alpha_1}} \, f_{S_{N-k},Y} (s,y) \, ds \, dy},
\end{equation}
where $Q_{\alpha}=Q_{\alpha}(S_{N-k}), Q_{\delta}=Q_{\delta}(Y), \alpha_1=\alpha+(1-\alpha)^{a+1}$ and $\delta_1=\delta+(1-\delta)^{d+1}$.

\bigskip
In practice, joint probability function is difficult to find unless a bivariate normal distribution is assumed. For the case of joint exponential distribution, we may refer to Kang et al. (2019) for Sarmanov's bivariate exponential distribution. In most cases, two or more dependent risks rely on Copula in order to have explicit formula of its joint distribution.

\bigskip \noindent \textsc{Proposition 2} \\
Let $S_{N-k}$ and $Y$ be two random losses with a joint distribution function represented by a Copula $C$. The Dependent Conditional VaR (DCoVaR) of $S_{N-k}$ given values beyond its VaR up to a fixed value of losses and a random loss $Y$ is given by
\begin{equation}
{\rm DCoVaR}_{(\alpha,a)}^{(\delta,d)}(S_{N-k}|Y;C) = \frac{\int\limits_{Q_{\alpha}}^{Q_{\alpha_1}} \, \int\limits_{Q_{\delta}}^{Q_{\delta_1}} \, s \, c(F_{S_{N-k}}(s),F_Y(y)) \, f_{S_{N-k}}(s) \, f_Y(y) \, dy \, ds}{C(\alpha_1,\delta_1) - C(\alpha,\delta_1) - C(\alpha_1,\delta) + C(\alpha,\delta)},
\end{equation}
where $F_{S_{N-k}}$ denote distribution function of $S_{N-k}$, $\alpha_1=\alpha+(1-\alpha)^{a+1}$ and $\delta_1=\delta+(1-\delta)^{d+1}$.	

\bigskip
\noindent \textsc{Remark.} According to the method of Brahim et al. (2018), the DCoVaR formula is represented by
\begin{equation}
{\rm DCoVaR}_{(\alpha,a)}^{(\delta,d)}(S_{N-k}|Y;C) = \frac{\int\limits_{\alpha}^{{\alpha_1}} \, \int\limits_{\delta}^{{\delta_1}} \, F_{S_{N-k}}^{-1}(u) \, c(u,v) \, dv \, du}{C(\alpha_1,\delta_1) - C(\alpha,\delta_1) - C(\alpha_1,\delta) + C(\alpha,\delta)},
\end{equation}
where $F_{S_{N-k}}^{-1}$ denote quantile function of $S_{N-k}$, $u=F_{S_{N-k}}(s), v=F_Y(y),$ $\alpha_1=\alpha+(1-\alpha)^{a+1}$ and $\delta_1=\delta+(1-\delta)^{d+1}$.	This formula, however, may not be obtained when no closed form expression of the quantile function is given.

\bigskip
The following properties apply to our proposed DCoVaR. The first property is to argue that the DCoVaR satisfies coherent property of risk measure in particular the subadditivity i.e. the DCoVaR of aggregate risk is no more than aggregate of DCoVaR of individual risk. Meanwhile, the second property is to show that the DCoVaR outperforms than MCoVaR and CCoVaR.

\bigskip \noindent \textsc{Property 1.} The Dependent Conditional VaR (DCoVaR) is a coherent risk measure.

\bigskip \noindent \textsc{Property 2.} The Dependent Conditional VaR (DCoVaR) has larger risk than or equal to MCoVaR and lower risk than or equal to CCoVaR.

\section{DCoVaR forecast for Pareto random loss}

Suppose that $X_i$, component for aggregate risk $S_{N-k}$, is a Pareto random loss with parameter $(1,\beta_i)$. We consider a dependent random loss $Y$ that follows a Pareto distribution with parameter $(1,\beta_a)$. The distribution functions of $X_i$ and $Y$ are, respectively, $F_{X_i}(x)=1-(\beta_i/(x+\beta_i))$, for $x_i \ge 0$, and $F_Y(y)=1-(\beta_a/(y+\beta_a))$, for $y \ge 0$. Their inverses are easy to find and thus their VaR's are straightforward i.e. ${\rm VaR}_{\alpha}(X_i) = Q_{\alpha} = \beta_i \, \big[ (1-\alpha)^{-1}-1 \big]$. In what follows, we provide some examples.

\bigskip \noindent \textsc{Example-1: DCoVaR forecast of a Pareto risk with a Pareto marginal.} The risk measure of DCoVaR forecast for $S_1=X_i$, given $Y$, may be found by using Proposition 2 since we apply a Copula for their distribution function. Specifically, we employ the Farlie-Gumbel-Morgenstern (FGM): $C_{\theta}^{{\rm FGM}}(u,v) = u \, v + \theta \, u \, v \, (1-u) \, (1-v)$, where $u,v \in [0,1], \theta \in [-1,1]$. Suppose that the joint distribution of $S_1$ and $Y$, defined by an FGM copula, is $F_{S_1,Y}(s,x)=C_{\theta_{SY}}(F_{S_1}(s),F_{Y}(y))$, where  $\theta \in [-1,1].$ Then, the DCoVaR of $S_1$ at levels of $\alpha$ and $\delta,\, 0<\alpha,\delta<1,$ is given by
\begin{align}
&{\rm DCoVaR}_{\left( {\alpha,a} \right)}^{\left( {\delta,d} \right)}\left( {{S_1}|Y;C} \right) = \frac{\beta_1 \, (\delta_1-\delta)}{{C\left( {{\alpha_1},{\delta_1}} \right) - C\left( {\alpha,{\delta_1}} \right) - C\left( {{\alpha_1},\delta } \right) + C\left( {\alpha,\delta } \right)}} \nonumber \\
&\qquad \times \Big\{ \Big[ \alpha_1-\alpha -\ln (1-(1-\alpha)^a) \Big] \Big[ 1+\theta_{SY} (1-\delta_1 -\delta) \Big] \nonumber \\
&\qquad \quad -\theta_{SY} (1-\delta_1 -\delta) \Big[ \alpha (\alpha-2) -\alpha_1 (\alpha_1-2) - \frac{1}{2} \ln \big( 1-(1-\alpha)^{\alpha} \big) \Big] \Big\}
\end{align}
where the Copulas are $C\left( {\alpha_1,\delta_1 } \right) = \alpha_1 \delta_1  + \theta_i \alpha_1 \delta_1 \left( {1 - \alpha_1 } \right)\left( {1 - \delta_1 } \right)$, $C\left( {\alpha,\delta_1 } \right) = \alpha \delta_1  + \theta_i \alpha \delta_1 \left( {1 - \alpha } \right)\left( {1 - \delta_1 } \right), C\left( {\alpha_1,\delta } \right) = \alpha_1 \delta  + \theta_i \alpha_1 \delta \left( {1 - \alpha_1 } \right)\left( {1 - \delta } \right)$, and $C\left( {\alpha,\delta } \right) = \alpha \delta  + \theta_i \alpha \delta \left( {1 - \alpha } \right)\left( {1 - \delta } \right)$. Meanwhile, when applying the method of Brahim et al., we find the DCoVaR as follows
\begin{align}
&{\rm DCoVaR}_{\left( {\alpha,a} \right)}^{\left( {\delta,d} \right)} \left( {S_1}|Y;C \right) = \frac{\beta_1(\delta_1-\delta)}{{C\left( {{\alpha_1},{\delta_1}} \right) - C\left( {\alpha,{\delta_1}} \right) - C\left( {{\alpha_1},\delta } \right) + C\left( {\alpha,\delta } \right)}} \nonumber \\
&\qquad \times \Big\{ \Big[ \ln (1-(1-\alpha)^a) \Big] \Big[ \theta_{SY} (1-\delta_1 -\delta)-1 \Big] \nonumber \\
&\qquad \quad - \Big[ \theta_{SY} \, (1+\alpha_1-\alpha)(1-\delta_1-\delta)-1 \Big]  (\alpha_1-\alpha) \Big\}
\end{align}

\bigskip \noindent \textsc{Example-2.} DCoVaR of Pareto risk in Example-1 may be carried out by using a Clayton Copula (which is an Archimedean Copula): $C^\text{C}_{\theta}(u,v) = \big( u^{-\theta} + v^{-\theta} - 1 \big)^{-1/\theta}$. The resulting DCoVaR forecast, however, is not in a closed form expression.
\begin{align}
&\text{DCoVaR}_{(\alpha,a)}^{(\delta,d)}(S_1|Y;C) = \frac{{{\beta _i}}}{{C\left( {{\alpha _1},{\delta _1}} \right) - C\left( {\alpha ,{\delta _1}} \right) - C\left( {{\alpha _1},\delta } \right) + C\left( {\alpha ,\delta } \right)}} \nonumber\\
&\quad\ \ \ \times \Bigg[ {\int\limits_\alpha ^{{\alpha _1}} {\frac{{{{\left( {{u^{ - \theta }} + \delta _1^{ - \theta } - 1} \right)}^{ - \frac{{1 + \theta }}{\theta }}} - {{\left( {{u^{ - \theta }} + {\delta ^{ - \theta }} - 1} \right)}^{ - \frac{{1 + \theta }}{\theta }}}}}{{{{\left( {1 - u} \right)}}{u^{\theta  + 1}}}} \, du} } \nonumber \\
&\qquad \quad - \Big\{ {{{\left( {\alpha _1^{ - \theta } + \delta _1^{ - \theta } - 1} \right)}^{ - \frac{{1 + \theta }}{\theta }}} - {{\left( {{\alpha ^{ - \theta }} + \delta _1^{ - \theta } - 1} \right)}^{ - \frac{{1 + \theta }}{\theta }}}} \Big\} \nonumber\\
&\qquad \quad + \Big\{ {{{\left( {\alpha _1^{ - \theta } + {\delta ^{ - \theta }} - 1} \right)}^{ - \frac{{1 + \theta }}{\theta }}} - {{\left( {{\alpha ^{ - \theta }} + {\delta ^{ - \theta }} - 1} \right)}^{ - \frac{{1 + \theta }}{\theta }}}} \Big\} \Bigg].
\end{align}

\bigskip
\noindent \textsc{Example-3.} DCoVaR for multivariate risk forecast may be expressed for the case of  $N$ identical dependent Pareto random risks: $X_i,\cdots,X_N$. Their joint probability function is given by
\[ f(x_1,\cdots,x_n;\gamma,N) = \frac{\Gamma(\gamma+N)}{\Gamma(\gamma)\beta^N} \, \frac{1}{\left(1+\frac{1}{\beta} \sum_{i=1}^{N} x_i\right)^{\gamma+N}}. \]
Let $S_N=X_1+\cdots+X_N$ and $Y$ be another Pareto random  risk with parameter $(1,\beta_a)$. Suppose that the joint distribution of $S_N$ and $Y$ is defined by a bivariate FGM Copula $F_{S_N,Y}(s,y) = C_{\theta_{SY}}(F_{S_N}(s),F_{Y}(y))$, where $\theta \in [-1,1]$. Then, for $N$ even, the DCoVaR of $S_N$ at levels $\alpha$ and $\delta,\, 0<\alpha,\delta<1,$ is given by
\begin{align}
&\text{DCoVaR}_{(\alpha,a)}^{(\delta,d)} \Big( S_N|Y;C \Big) = \frac{N \, \gamma(\delta_1-\delta)}{{C\left( {{\alpha_1},{\delta_1}} \right) - C\left( {\alpha,{\delta_1}} \right) - C\left( {{\alpha_1},\delta } \right) + C\left( {\alpha,\delta } \right)}} \nonumber\\
&\times \Bigg\{ \Big( \delta_1-\delta \Big)  \Big( 1+\theta_{SY}(1-\delta_1-\delta) \Big) \times \Bigg[ \ln \frac{1-\alpha^{1/N}}{1-\alpha^{1/N}}+N\left(\alpha^{1/N}-\alpha_1^{1/N} \right)-\cdots \nonumber\\
&\qquad + \frac{N}{N-1} \left[ \left(1-\alpha^{1/N} \right)^{N-1} -\left(1-\alpha_1^{1/N} \right)^{N-1} \right] - \frac{1}{N} \left[ \left(1-\alpha^{1/N} \right)^{N} -\left(1-\alpha_1^{1/N} \right)^{N} \right] \Bigg]  \nonumber \\
&\qquad + 2 \, \theta_{SY} \, (1-\delta_1-\delta) \times \Bigg[ \ln \frac{1-\alpha^{1/N}}{1-\alpha^{1/N}} + N \left(\alpha^{1/N}-\alpha_1^{1/N} \right) - \cdots \nonumber\\
&\qquad + \frac{2N}{2N-1} \left[ \left(1-\alpha^{1/N} \right)^{2N-1} -\left(1-\alpha_1^{1/N} \right)^{2N-1} \right] - \frac{1}{2N} \left[ \left(1-\alpha^{1/N} \right)^{2N} - \left(1-\alpha_1^{1/N} \right)^{2N} \right] \Bigg] \Bigg\}
\end{align}
whilst for $N$ odd, the DCoVaR of $S_N$ at levels $\alpha$ and $\delta,\, 0<\alpha,\delta<1,$ is given by

\begin{align}
&\text{DCoVaR}_{(\alpha,a)}^{(\delta,d)} \Big( S_N|Y;C \Big) = \frac{N\gamma(\delta_1-\delta)}{{C\left( {{\alpha_1},{\delta_1}} \right) - C\left( {\alpha,{\delta_1}} \right) - C\left( {{\alpha_1},\delta } \right) + C\left( {\alpha,\delta } \right)}} \nonumber\\
&\times \Bigg\{ \Big( \delta_1-\delta \Big) \Big( 1+\theta_{SY}(1-\delta_1-\delta) \Big) \times \Bigg[ \ln \frac{1-\alpha^{1/N}}{1-\alpha^{1/N}} + N\left(\alpha^{1/N}-\alpha_1^{1/N} \right)-\cdots \nonumber\\
&\qquad - \frac{N}{N-1} \left[ \Big(1-\alpha^{1/N} \Big)^{N-1} -\Big( 1-\alpha_1^{1/N} \Big)^{N-1} \right] + \frac{1}{N} \left[ \Big(1-\alpha^{1/N} \Big)^{N} - \Big(1-\alpha_1^{1/N} \Big)^{N} \right] \Bigg]  \nonumber \\
&\qquad + 2 \, \theta_{SY} \, \Big( 1-\delta_1-\delta \Big) \times \Bigg[ \ln \frac{1-\alpha^{1/N}}{1-\alpha^{1/N}} + N \Big( \alpha^{1/N}-\alpha_1^{1/N} \Big) - \cdots \nonumber\\
&\qquad + \frac{2N}{2N-1} \left[ \Big( 1-\alpha^{1/N} \Big)^{2N-1} - \Big( 1-\alpha_1^{1/N} \Big)^{2N-1} \right]  - \frac{1}{2N} \left[ \Big(1-\alpha^{1/N} \Big)^{2N} -\Big( 1-\alpha_1^{1/N} \Big)^{2N} \right] \Bigg] \Bigg\}
\end{align}

\bigskip \noindent \textsc{DCoVaR forecast for Pareto random loss: A simulation result} \\
We carry out a simulation study for calculating DCoVaR forecast. The parameters of Pareto distribution of $X_1$ and $Y$ are, respectively, $\beta_1=1.5$ and $\beta_a=1.5$. Suppose also the model parameter $\Lambda$ is gamma distributed with shape and scale parameters $\tau=\omega=1$. The significance level for $\alpha$ (and $\delta$) is set above 0.9 whilst we set $a=d=0.1$. Figure 1-3 show the DCoVaR forecast for the above parameters set up. As for comparison, we also plot the MCoVaR forecast. For each figure, we have an associate or dependent random loss $Y$ which is a Pareto random loss, an aggregate $Y=S_2$ of Pareto losses, and a parameter model $Y=\Lambda$ of gamma distributed. It is shown from the figures that the DCoVaR forecast tends to increase as $\delta$ increases whilst the MCoVaR forecast remains the same. As for the CCoVaR forecast, it is larger than the DCoVaR forecast (not shown in the figures).

\bigskip
\begin{figure}[htpb!]
\begin{center}	
\includegraphics[scale=0.3]{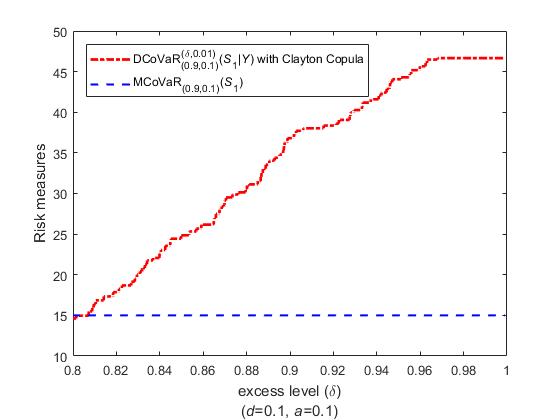}
\includegraphics[scale=0.3]{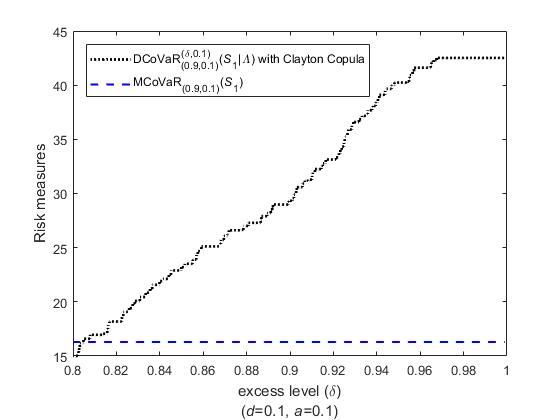}
\includegraphics[scale=0.3]{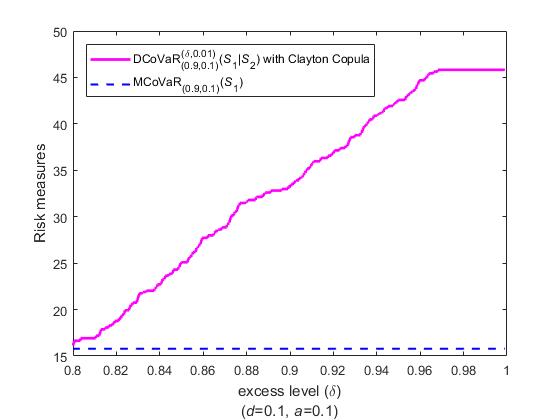}
\caption{DCoVaR forecast of $S_1$ with various of $Y$ and using Clayton Copula; $Y$ is Pareto distributed, $Y=S_2$ is Pareto distributed, $Y=\Lambda$ is Gamma distributed; such forecasts are in comparison to MCoVaR forecast.}
\label{figure2}
\end{center}
\end{figure}

\begin{figure}[htpb!]
\begin{center}	
\includegraphics[scale=0.3]{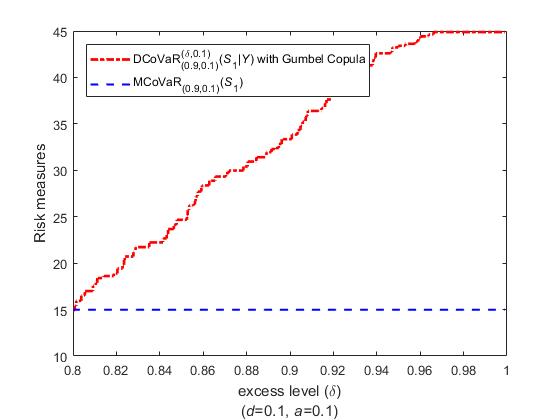}
\includegraphics[scale=0.3]{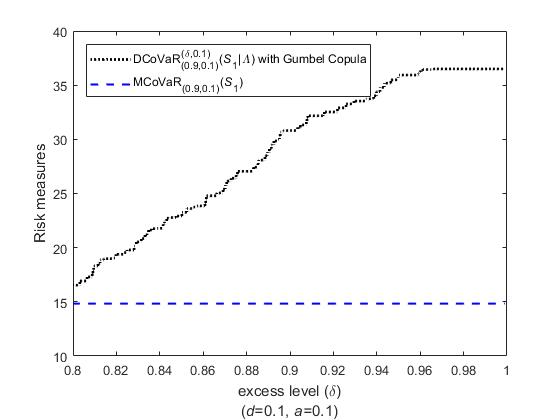}
\includegraphics[scale=0.3]{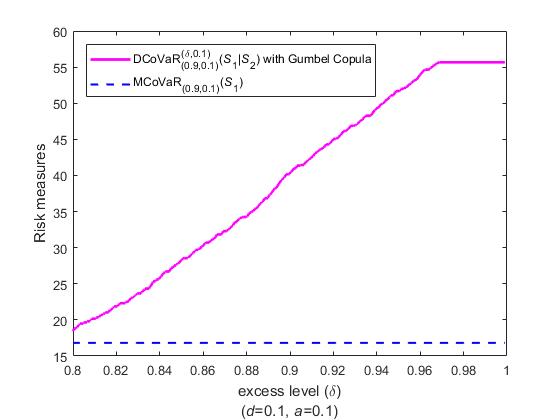}
\caption{DCoVaR forecast of $S_1$ with various of $Y$ and using Gumbel Copula; $Y$ is Pareto distributed, $Y=S_2$ is Pareto distributed, $Y=\Lambda$ is Gamma distributed; such forecasts are in comparison to MCoVaR forecast}
\label{figure2}
\end{center}
\end{figure}

\begin{figure}[htpb!]
\begin{center}	
\includegraphics[scale=0.3]{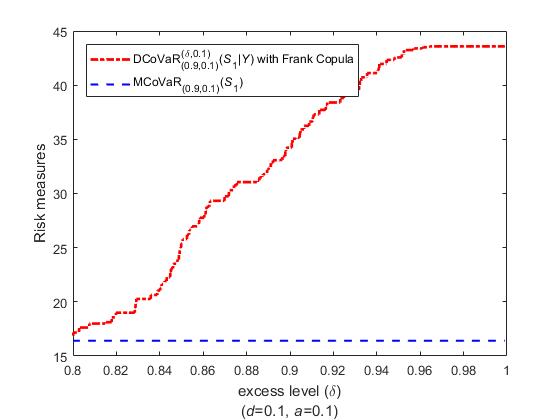}
\includegraphics[scale=0.3]{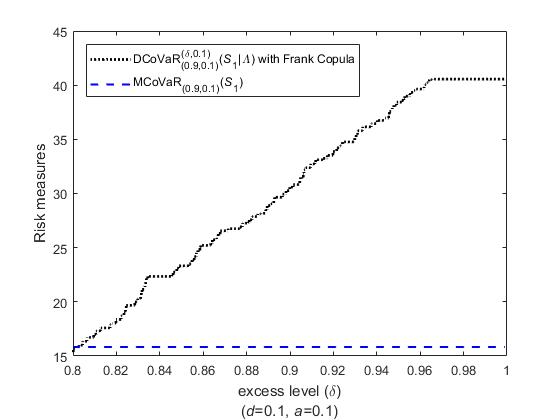}
\includegraphics[scale=0.3]{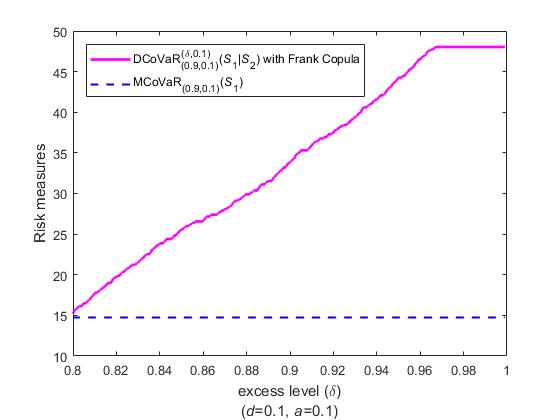}
\caption{DCoVaR forecast of $S_1$ with various of $Y$ and using Frank Copula; $Y$ is Pareto distributed, $Y=S_2$ is Pareto distributed, $Y=\Lambda$ is Gamma distributed; such forecasts are in comparison to MCoVaR forecast}
\label{figure2}
\end{center}
\end{figure}

Note that, as for the Copula choices, we have used Archimedean Copulas. The Clayton Copula (Figure 1) function is given by $C_\theta^C(u,v) = \Big( u^{-\theta}+v^{-\theta}-1 \Big)^{-1/\theta}, \theta \in [-1,\infty)$. Meanwhile, for other Copulas of Gumbel (Figure 2) and Frank (Figure 3) the functions are $C_\theta^G(u,v) = \exp \Big\{- \Big[ (-\ln u)^\theta+(-\ln u)^\theta \Big]^{1/\theta} \Big\}, \theta \in [0,\infty)$ and $C_\theta^F(u,v) = -\frac{1}{\theta} \, \ln \Big( 1 - \frac{(1-e^{-\theta u})(1-e^{-\theta v})}{1-e^{-\theta}} \Big), \theta \in (-\infty ,0) \cup (0,\infty)$, respectively.

The significance level for $\alpha$ (and $\delta$) is set above 0.9. Unlike calculating CoVaR forecast, the DCoVaR forecast computation requires two significance levels. In particular, the joint significance level is given by
\[ P\Big( Q_\alpha \le S_1 \le Q_{\alpha_1},Q_\delta \le Y \le Q_{\delta_1} \Big) = C_\theta(\alpha_1,\delta_1) - C_\theta(\alpha,\delta_1) - C_\theta(\alpha_1,\delta) + C_\theta(\alpha,\delta). \]
For the case of $a=d=0$, the joint significance level is $1 - \alpha - \delta + C_\theta(\alpha,\delta)$. We use joint significance level to measure the number of violations of the DCoVaR forecast. We generate data of 3000 observations for each $X_1$, $Y$, and $\Lambda$. The DCoVaR forecast is computed by using Proposition 2.

Assessment of accuracy for the DCoVaR forecast is carried out by first observing joint significance level. For example, in Table 1 (first row, first column), 2.79\% joint significance level is lower than 10\%. This means that the DCoVaR forecast is quite accurate. Then, by calculating the number of violations of the DCoVaR$_{(\alpha,0.1)}^{(\delta,0.1)}$, it is obtained 1.83\% (number of violations is 55, total observations 3000; 55/3000=0.0183). Basically, the numbers of violations are the number of sample observations located out of the critical value i.e. more than or equal to DCoVaR$_{(\alpha,0.1)}^{(\delta,0.1)}$ forecast. These computations are shown in Table 1-3, for Clayton, Gumbel, and Frank Copulas, respectively. In summary, using Clayton Copula provides more accurate forecast due to lower joint significance level and number of violations.

\begin{table}[htpb!]
\caption{Joint significance level and number of violations of the DCoVaR$_{(\alpha,0.1)}^{(\delta,0.1)}$ forecast of $S_1$ associated with $Y$ with Clayton Copula $(\theta=7.0)$, $a=d=0.1$}
\label{correctClayton}
\begin{center}
\begin{tabular}{cccccc}
\hline
& & $\alpha=0.90$ & & $\alpha=0.95$ & \\
& & sig. level (\%) & no. violations & sig. level (\%) & no. violations \\
& & & (\%) & & (\%) \\
\hline
\hline
$\delta$ & 0.9000 & 2.79 & 55 & 1.43 & 35 \\
& & & (1.83) & & (1.17) \\
& 0.9250 & 2.13 & 38 & 1.12 & 24 \\
& & & (1.27) & & (0.80) \\
& 0.9500 & 1.43 & 22 & 0.77 & 12 \\
& & & (0.73) & & (0.40) \\
\hline
\end{tabular}
\end{center}
\end{table}

\begin{table}[htpb!]
\caption{Joint significance level and number of violations of the DCoVaR$_{(\alpha,0.1)}^{(\delta,0.1)}$ forecast of $S_1$ associated with $Y$ with Gumbel Copula $(\theta=6.3)$, $a=d=0.1$}
\label{correctGumbel}
\begin{center}
\begin{tabular}{cccccc}
\hline
& & $\alpha=0.90$ & & $\alpha=0.95$ & \\
& & sig. level (\%) & no. violations & sig. level (\%) & no. violations \\
& & & (\%) & & (\%) \\
\hline
\hline
$\delta$ & 0.9000 & 6.61 & 140 & 2.91 & 83 \\
& & & (4.67) & & (2.77) \\
& 0.9250 & 5.14 & 119 & 3.17 & 80 \\
& & & (3.97) & & (2.67) \\
& 0.9500 & 2.91 & 78 & 2.99 & 62 \\
& & & (2.60) & & (2.07) \\
\hline
\end{tabular}
\end{center}
\end{table}

\begin{table}[htpb!]
\caption{Joint significance level and number of violations of the DCoVaR$_{(\alpha,0.1)}^{(\delta,0.1)}$ forecast of $S_1$ associated with $Y$ with Frank Copula $(\theta=25)$, $a=d=0.1$}
\label{correctFrank}
\begin{center}
\begin{tabular}{cccccc}
\hline
& & $\alpha=0.90$ & & $\alpha=0.95$ & \\
& & sig. level (\%) & no. violations & sig. level (\%) & no. violations \\
& & & (\%) & & (\%) \\
\hline
\hline
$\delta$ & 0.9000 & 4.42 & 101 & 2.21 & 54 \\
& & & (3.37) & & (1.80) \\
& 0.9250 & 3.41 & 80 & 1.91 & 52 \\
& & & (2.67) & & (1.73) \\
& 0.9500 & 2.21 & 50 & 1.41 & 32 \\
& & & (1.67) & & (1.07) \\
\hline
\end{tabular}
\end{center}
\end{table}

\section{Application to financial returns data}

We carry out a numerical analysis of returns data and model it with stochastic volatility processes. In particular, we employ the Generalized Autoregressive Conditional Heteroscedastic (GARCH) model of order one. Consider two returns processes, $\{ X_{1t} \}$ and $\{ X_{2t} \}$. Suppose that each process follows a GARCH(1,1) model defined as
\[ X_{it} = \varepsilon_t \, \sqrt{h_t}, \, \, \, h_t = \kappa_0 + \kappa_1 \, X_{t-1}^2 + \eta \, h_{t-1}, \, \, \, i=1,2 \]
where $\kappa_0>0, \kappa_1 \ge 0, \eta \ge 0,$ and $\kappa_1 + \eta<1$. Let $S_t=X_{1t}$ and $Y_t=X_{2t}$. The DCoVaR forecast of $S_t$ with an associate risk $Y_t$ is given by
\begin{align}
&\text{DCoVaR}_{(\alpha,a)}^{(\delta,d)}(S_t|Y_t;C) = E(S_t|Q_t^\alpha < S_t <Q_t^{\alpha_1},Q_t^\delta < Y_t <Q_t^{\delta_1})\nonumber\\
&\qquad = \frac{\int\limits_{Q_t^\alpha}^{Q_t^{\alpha_1}} \int\limits_{Q_t^\delta}^{Q_t^{\delta_1}} s_t\, c(F(s_t),F(y_t)|\mathcal{G}_{t-1}) f(s_t|\mathcal{G}_{t-1})f(y_t|\mathcal{G}_{t-1})\, dy_t \, ds_t}{P(Q_t^\alpha < S_t <Q_t^{\alpha_1},Q_t^\delta < Y_t <Q_t^{\delta_1})} \label{1dcovar11}
\end{align}
where $f_{S_{t}}(\cdot|\mathcal{G}_{t-1})$ is the conditional probability function of the target risk $S_{t}$ on $\mathcal{G}_{t-1}$. The denominator of (\ref{1dcovar11}) is given by
\[ P\Big( Q_t^\alpha < S_t <Q_t^{\alpha_1},Q_t^\delta < Y_t <Q_t^{\delta_1} \Big) = \int_{Q_t^\alpha}^{Q_t^{\alpha_1}} \int_{Q_t^\delta}^{Q_t^{\delta_1}} c(F(s_t),F(y_t)|\mathcal{G}_{t-1}) f(s_t|\mathcal{G}_{t-1})f(y_t|\mathcal{G}_{t-1})\, dy_t \, ds_t, \]
and $Q_t^\alpha$ as well as $Q_t^\delta$ satisfy
\[ P(S_t \le Q_t^\alpha|\mathcal{G}_{t-1}) = \int\limits_{-\infty}^{Q_t^\alpha} f(s_t|\mathcal{G}_{t-1}) \, ds_t, \, \, \, P(Y_t \le Q_t^\delta|\mathcal{G}_{t-1}) = \int\limits_{-\infty}^{Q_t^\delta} f(y_t|\mathcal{G}_{t-1}) \, dy_t. \]

\begin{figure}[htpb!]
\begin{center}
\centering
\includegraphics[scale=0.6]{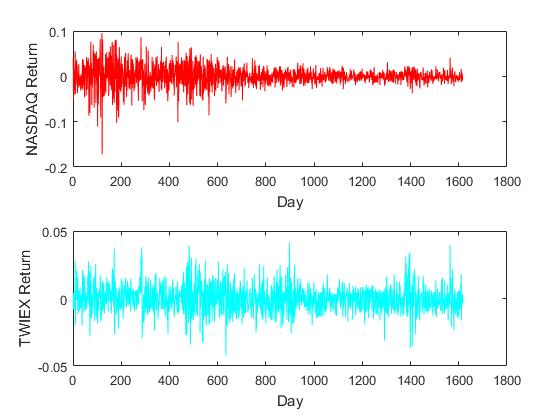}
\caption{Daily returns of NASDAQ and TWIEX. A stylized fact of volatility clustering may be observed for both returns.}
\label{gambar1}
\end{center}
\end{figure}

\bigskip \noindent \textsc{Empirical results} \\
We have used the data of NASDAQ and TWIEX assets from July 3, 2000 to May 17, 2007, taken from www.yahoofinance.com for total of 1617 observations. We define loss data as the negative return of an asset formulated as follows
\[ X_{it} = - \ln \left( \frac{P_{it}}{P_{i,t-1}} \right), \]
where $P_{it}$ is the price of an $i$-th asset at time $t$, $i=1,2$. Figure \ref{gambar1} shows such daily returns. In addition, we may observe that one of the stylized facts of returns, known as volatility clustering, occurs in both NASDAQ and TWIEX returns. Huang et al. (2009) argued that the GARCH-$t$(1,1) model were appropriate for the returns of NASDAQ and TWIEX. Accordingly, we presents the maximum likelihood estimates for such model parameter as in Table \ref{garch}.

\begin{table}[htpb!]
\caption{Parameters estimates of GARCH-$t$(1,1) model for NASDAQ and TWIEX.}
\label{garch}
\begin{center}
\begin{tabular}{ccccc}
\hline
& $\widehat{\kappa}_0$ & $\widehat{\kappa}_1$ & $\widehat{\eta}_1$ & $\widehat{\nu}$ \\
\hline
\hline
NASDAQ & 0.0064 & 0.0266 & 0.9678 & 6.4188 \\
TWIEX & 0.0368 & 0.0643 & 0.9082 & 6.9057 \\
\hline
\end{tabular}
\end{center}
\end{table}

\begin{figure}[htpb!]
\begin{center}
\centering
\includegraphics[scale=0.6]{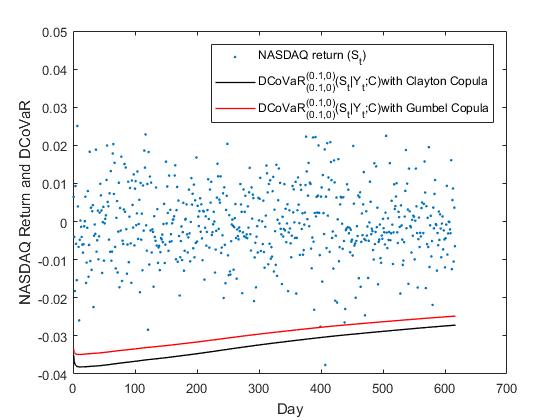}
\caption{DCoVaR forecast of the NASDAQ returns, given the TWIEX returns.}
\label{fig5}
\end{center}
\end{figure}

\begin{table}[htpb!]
\caption{Joint significance level (\%) and number of violations (\%) of the DCoVaR$_{(\alpha,0)}^{(\delta,0)}$ forecast by using Clayton and Gumbel Copulas.}
\label{correct2}
\begin{center}
\begin{tabular}{ccccc}
\hline
Copula & Parameter & & & \\
Clayton & $\alpha=\delta=0.10$ & $\alpha=\delta=0.15$ & $\alpha=0.10,\delta=0.15$ & $\alpha=0.15,\delta=0.10$ \\
($\widehat{\theta}=0.4938$) & $C_\theta(0.10,0.10)$ & $C_\theta(0.15,0.15)$ & $C_\theta(0.10,0.15)$ & $C_\theta(0.15,0.10)$ \\
\hline
Joint sig. level & 3.49 & 5.72 & 4.41 & 4.41 \\
\hline
No. violations & 0.16 & 0.65 & 0.16 & 0.65 \\
\hline
\hline
Copula & Parameter & & & \\
Gumbel & $\alpha=\delta=0.10$ & $\alpha=\delta=0.15$ & $\alpha=0.10,\delta=0.15$ & $\alpha=0.15,\delta=0.10$ \\
($\widehat{\theta}=1.2905$) & $C_\theta(0.10,0.10)$ & $C_\theta(0.15,0.15)$ & $C_\theta(0.10,0.15)$ & $C_\theta(0.15,0.10)$ \\
\hline
Joint sig. level & 1.95 & 3.90 & 2.74 & 2.74 \\
\hline
No. violations & 0.16 & 0.49 & 0.16 & 0.49 \\
\hline
\end{tabular}
\end{center}
\end{table}

In order to calculate the DCoVaR forecast, Figure \ref{fig5}, we do in-sample forecast in which we have used 1000 first observations whilst the out-of-sample is to evaluate forecasting performance. As in Table \ref{garch} above, Student's $t$ distribution is assumed for innovation. Meanwhile, Archimedean Copula are used for the joint distribution function. In particular, we employ Clayton and Gumbel Copulas. The parameter $\theta$ for each Copula is estimated by maximum likelihood method. We obtain $\widehat{\theta}_C=0.4938$ and $\widehat{\theta}_G=1.2905$, respectively.

The number of violations of the DCoVaR$_{(\alpha,0)}^{(\delta,0)}$ forecast for both Clayton and Gumbel Copulas are presented in Table \ref{correct2}. It is the number of sample
observations located out of the critical value i.e. less than or equal the DCoVaR forecast. It is shown from the table that the DCoVaR forecast with Gumbel Copula has lower joint significance level in comparison to the DCoVaR forecast with Clayton Copula. As for the number of violations, it conforms  the use of Gumbel Copula. In short, it suggests that Gumbel Copula is more appropriate Copula for describing the joint distribution of NASDAQ and TWIEX returns.

\section{Concluding remark}

The use of GARCH model for marginal of asset returns may be replaced by its extensions such as ARMA-GARCH and GJR-GARCH models. In addition, any innovations may also be applied to such volatility models. Syuhada (2020) has carried out VaR forecast and compared such observable stochastic volatility process (GARCH) class of models) with the latent one i.e the Stochastic Volatility Autoregressive (SVAR) model.

\section*{Acknowledgment}

We are grateful to anonymous referees for their comments that improve the paper. We also thank Prof Ken Seng Tan (University of Waterloo) for a thoughtful discussion.

\newpage
\section*{References}

\begin{enumerate}
\item Acerbi, C., and D. Tasche. 2002. On the coherence of expected shortfall. {\it Journal of Banking and Finance} 26(7):1487-1503.

\item Brahim, B., Fatah, B., and Y. Djabrane. 2018. Copula conditional tail expectation for multivariate financial risks. {\it Arab Journal of Mathematical Sciences} 24(1):82-100.

\item Huang, J., Lee, K., Liang, H., and W. Lin. 2009. Estimating value-at-risk of portfolio by conditional copula-GARCH model. {\it Insurance: Mathematics and Economics} 45:315-324.

\item Jadhav, D., Ramanathan, T., and U. Naik-Nimbalkar. 2009. Modified estimators of the expected shortfall. {\it Journal of Emerging Market Finance} 8(2):87:107.

\item Jadhav, D., Ramanathan, T., and U. Naik-Nimbalkar. 2013. Modified expected shortfall: A new robust coherent risk measure. {\it Journal of Risk} 16(1):69-83.

\item Kabaila P., and K. Syuhada. 2008. Improved prediction limits for AR($p$) and ARCH($p$) processes. {\it Journal of Time Series Analysis} 29:213-223.

\item Kabaila P., and K. Syuhada. 2010. The asymptotic efficiency of improved prediction intervals. {\it Statistics and Probability Letters} 80(17-18):1348-1353.

\item Kabaila, P. and R. Mainzer. 2018. Estimation risk for value-at-risk and expected shortfall. {\it Journal of Risk} 20(3):29-47.

\item Kang, Y., Wang, D., and J. Cheng. 2019. Risk models based on copulas for premiums and claim sizes. {\it Communications in Statistics - Theory and Methods}. doi:10.1080/03610926.2019.1662443.

\item Koji, I., and M. Kijima. 2003. On the significance of expected shortfall as a coherent risk measure {\it Journal of Banking \& Finance} 24(9): 853-864.

\item McNeil, A., Frey, R., and P. Embrechts. 2005. {\it Quantitative Risk Management: Concepts, Techniques, and Tools}, Princeton University Press.

\item Nadarajah, S., Chan, S., and E. Afuecheta. 2016. Tabulations for value at risk and expected shortfall. {\it Communications in Statistics - Theory and Methods}. doi:10.1080 /03610926.2015.1116572.

\item Nieto, M.R., and E. Ruiz. 2016. Frontiers in VaR forecasting and backtesting. {\it International Journal of Forecasting} 32: 475-501.

\item Syuhada, K. 2020. The improved Value-at-Risk for heteroscedastic processes and their coverage probability. {\it Journal of Probability and Statistics} Article ID 7638517.

\item Syuhada, K., Nur'aini, R., and Mahfudhotin. 2020. Quantile-based estimative VaR forecast and dependence measure: A simulation approach. {\it Journal of Applied Mathematics} Article ID 8276019.

\item Yan, Z., and J. Zhang. 2016. Adjusted Empirical Likelihood for Value at Risk and Expected Shortfall. {\it Communications in Statistics - Theory and Methods}. doi:10.1080/03610926.2014.1002933.

\item Zhang, Y., Zhao, P., and K. Cheung. 2018. Comparison of aggregate claim numbers and amounts: A study of heterogeneity. {\it Scandinavian Actuarial Journal}. doi:10.1080/03461238.2018.1557738.
\end{enumerate}

\newpage
\section*{Appendix}

\bigskip \noindent \textsc{Proof for Proposition 1.} For simplicity, let
\begin{align*}
Q_p = Q_p(S_{N-k}),\quad Q_{p_1}= Q_{p_1}(S_{N-k}),\quad Q_\delta = Q_\delta (Y),\quad \text{and} \quad Q_{\delta_1} = Q_{\delta_1}(Y).
\end{align*}
Then	
\begin{align*}
&\text{DCoVaR}_{(\alpha,a)}^{(\delta,d)} \Big( S_{N-k} \, \Big| \, Y \Big) \\
&\qquad = \frac{1}{\text{P}(Q_\alpha \le S_{N-k} \le Q_{\alpha_1}, Q_\delta \le S_{N-k} \le Q_{\delta_1})} \times E[S_{N-k}\,\textbf{1}_{(Q_\alpha \le S_{N-k} \le Q_{\alpha_1}, Q_\delta \le S_{N-k} \le Q_{\delta_1})}]\\
&\qquad = \frac{\int\limits_{Q_{\delta}}^{Q_{\delta_1}} \, \int\limits_{Q_{\alpha}}^{Q_{\alpha_1}} \, s \, f_{S_{N-k},Y} (s,y) \, ds \, dy}{\int\limits_{Q_{\delta}}^{Q_{\delta_1}} \, \int\limits_{Q_{\alpha^1}}^{Q_{\alpha_1}} \, f_{S_{N-k},Y} (s,y) \, ds \, dy}
\end{align*}

\bigskip
\bigskip \noindent \textsc{Proof for Proposition 2.} We assume first that $s \le Q_{p_1}(S_{N-k})$.
We obtain
\begin{align*}
&\text{P}(S_{N-k} \le s|Q_p \le S_{N-k} \le Q_{p_1},Q_\delta \le Y \le Q_{\delta _1}) = \frac{\text{P}(Q_p \le S_{N-k} \le s,Q_\delta \le Y \le Q_{\delta _1})}{\text{P}(Q_p \le S_{N-k} \le Q_{p_1},Q_\delta \le Y \le Q_{\delta _1})},
\end{align*}
where the denominator may be written as
\begin{align*}
&\text{P}(Q_p \le S_{N-k} \le Q_{p_1},Q_\delta \le Y \le Q_{\delta _1}) = C(p_1,\delta_1)- C(p,\delta_1)-C(p_1,\delta) + C(p,\delta).
\end{align*}	
Thus,
\begin{align*}
&\text{P}(S_{N-k} \le s|Q_p \le S_{N-k} \le Q_{p_1},Q_\delta \le Y \le Q_{\delta_1}) \\
&\qquad  = \frac{1}{C(p_1,\delta_1)-C(p,\delta_1)-C(p_1,\delta)+C(p,\delta)} \times \int\limits_{Q_\delta}^{Q_{\delta_1}} {\int\limits_{Q_p}^s \frac{{\partial ^2}C(F_{S_{N-k}}(s),F_Y(y))}{\partial s\,\partial y}ds\,dy}  .
\end{align*}
For fixed level $p=\alpha$ and $a$, the DCoVaR of $S_{N-k}$ is given by
\begin{align*}
&\text{DCoVaR}_{(\alpha,a)}^{(\delta,d)}(S_{N-k}|Y;C)\\
&\qquad = \frac{1}{C(p_1,\delta_1)-C(p,\delta_1)-C(p_1,\delta)+C(p,\delta)} \times \int\limits_{Q_\delta}^{Q_{\delta_1}} {\int\limits_{Q_\alpha}^{Q_{\alpha_1}} \frac{s\ \partial^2 C(F_{S_{N-k}}(s),F_Y(y ))}{\partial s\,\partial y}ds\,dy}  .
\end{align*}

\newpage
\noindent We suppose that the densities of $F_{S_{N-k}}$ and $F_Y$ are $f_{S_{N-k}}$ and $f_Y$, respectively. Thus, 
\begin{align*}
&\text{DCoVaR}_{(\alpha,a)}^{(\delta,d)}(S_{N-k}|Y;C)\\
&\qquad =\frac{1}{C(\alpha_1,\delta_1)-C(\alpha,\delta_1)-C(\alpha_1,\delta) + C(\alpha,\delta)} \times \int\limits_{Q_\delta}^{Q_{\delta_1}} {\int\limits_{Q_\alpha}^{Q_{\alpha_1}} s\,c(F_{S_{N-k}}(s),F_Y(y))}  \, f_{S_{N-k}}(s)f_Y(y)\,ds\,dy.
\end{align*}
	
\bigskip \noindent \textsc{Proof for Property 1.} To prove coherent property, we follow the proof of the subbaditivity of the CoVaR, given in Acerbi and Tasche (2002) and that of the MCoVaR, given in Jadhav et al. (2013). For simplicity, let $S_{N-k}=S.$
Let $F_S(s)$ be the distribution function of a continuous random variable $S$ and define the $\alpha$-quantile of $S$ as ${Q_\alpha}= F_{S}^{-1}( \alpha )$ for a specified probability $\alpha \in (0,1)$ and $\delta$-quantile of $Y$ as ${Q_\delta}= F_{Y}^{-1}( \delta)$ for some probability $\delta \in (0,1)$. We may write the DCoVaR as
\begin{align*}
&\text{DCoVaR}_{(\alpha,a)}^{(\delta,d)}(S|Y;C) \\
&\qquad= \frac{1}{C(\alpha_1,\delta_1)-C(\alpha,\delta_1)-C(\alpha_1,\delta) + C(\alpha,\delta)} \times E\Big[S\, \mathbf{1}_{\{Q_\alpha \le S \le {Q_{\alpha_1}},Q_\delta \le Y \le Q_{\delta_1} \}}\Big]. \end{align*}
	
\noindent Let $S_2 = S + S^1.$ Then
\begin{align*}
&(1-\delta)^{d+1} + C(\alpha,\delta)-C(\alpha,\delta_1)\\
&\qquad\times \Big\{\text{DCoVaR}_{(\alpha,a)}^{(\delta,d)}(S|Y;C) + \text{DCoVaR}_{( \alpha,a)}^{(\delta,d)}(S^1|Y;C) - \text{DCoVaR}_{(\alpha,a)}^{(\delta,d)}(S_2|Y;C)\Big\}\\
&\quad = E\Big[{S \Big(\mathbf{1}_{\{Q_\alpha^2 \le S_2 \le Q_{\alpha_1}^2,Q_\delta  \le Y \le Q_{\delta_1}\}} - \mathbf{1}_{\{Q_\alpha \le S \le Q_{\alpha_1},Q_\delta \le Y \le Q_{\delta_1}}\} \Big)}\\
&\qquad\quad+S^1 \Big( {\mathbf{1}_{\{Q_\alpha^2 \le S_2 \le Q_{\alpha_1}^2,Q_\delta  \le Y \le Q_{\delta_1}\}} - \mathbf{1}_{\{Q_\alpha \le S \le Q_{\alpha_1},Q_\delta \le Y \le Q_{\delta_1}\}}} \Big)\Big]\\
&\quad\ge Q_\alpha E\Big[\mathbf{1}_{\{Q_\alpha^2 \le S_2 \le Q_{\alpha_1}^2,Q_\delta \le Y \le Q_{\delta_1}\}} - \mathbf{1}_{\{Q_\alpha \le S \le Q_{\alpha_1},Q_\delta \le Y \le Q_{\delta _1}\}} \Big]\\
&\qquad\quad + Q_\alpha^1 E\Big[\mathbf{1}_{\{Q_\alpha^2 \le S_2 \le Q_{\alpha_1}^2,Q_\delta \le Y \le Q_{\delta_1}\}} - \mathbf{1}_{\{Q_\alpha^1 \le {S^1} \le Q_{\alpha_1}^1,Q_\delta \le Y \le Q_{\delta_1}\}} \Big]\\
&\quad = Q_\alpha \Big\{ C(\alpha_1,\delta_1)-C(\alpha,\delta_1)-C(\alpha_1,\delta)+C(\alpha,\delta) - C(\alpha_1,\delta_1)+C(\alpha ,\delta_1)+C(\alpha_1,\delta)-C(\alpha,\delta) \Big\} \\
&\qquad\quad + Q_\alpha^1 \Big\{ C(\alpha_1,\delta_1)-C(\alpha ,\delta_1)-C(\alpha_1,\delta)+C(\alpha,\delta) - C(\alpha_1,\delta_1)+C(\alpha ,\delta_1)+C(\alpha_1,\delta)-C(\alpha,\delta) \Big\} \\
&\qquad = 0.
\end{align*}
	
\noindent In the above inequality, we have used
\begin{enumerate}
\item[(*)] if ${S} <Q_\alpha$, then
    \[ \mathbf{1}_{\{Q_\alpha^2 \le S_2 \le Q_{\alpha_1}^2,Q_\delta \le Y \le Q_{\delta _1}\}} - \mathbf{1}_{\{Q_\alpha \le S \le Q_{\alpha_1},Q_\delta \le Y \le Q_{\delta_1} \}} \ge 0; \]
    
\item[(**)] if $Q_\alpha \le S \le Q_{\alpha_1}$, then
    \[ \mathbf{1}_{\{Q_\alpha^2 \le S_2 \le Q_{\alpha_1}^2,Q_\delta \le Y \le Q_{\delta _1}\}} - \mathbf{1}_{\{Q_\alpha \le S \le Q_{\alpha_1},Q_\delta \le Y \le Q_{\delta_1} \}} \le 0; \]
\end{enumerate}
This proves that the DCoVaR follows the subadditivity and hence is a coherent risk measure.

\bigskip \noindent \textsc{Proof for Property 2.} Note that the statement in Property 2 is mathematically equivalent to these both inequalities.
\begin{align*}
\text{MCoVaR}_{(\alpha,a)}(S)\le \text{DCoVaR}_{(\alpha ,a)}^{(\delta,0)}(S|Y;C), \\
\text{DCoVaR}_{(\alpha,a)}^{(\delta,0)}(S|Y;C) \le \text{CCoVaR}_\alpha^\delta(S|Y;C).
\end{align*}
Note that
\begin{enumerate}
\item We may write the MCoVaR as
\begin{align*}
\text{MCoVaR}_{(\alpha,a)}(S) = \frac{1}{(1-\alpha)^{a+1}}E[S\, \mathbf{1}_{\{Q_\alpha \le S \le Q_{\alpha_1}\}}].
\end{align*}
Thus,
\begin{align*}
&{(1-\alpha)^{a + 1}}\Big[\text{MCoVaR}_{(\alpha,a)}(S)-\text{DCoVaR}_{(\alpha,a)}^{(\delta,0)}(S|Y;C) \Big]\\
&\qquad= E\Big\{S\Big[\mathbf{1}_{\{Q_\alpha\le S \le Q_{\alpha_1}\}} - \frac{(1-\alpha)^{a+1}}{\alpha_1-\alpha-\delta+C(\alpha,\delta)}\mathbf{1}_{\{Q_\alpha \le S \le Q_{\alpha_1},Q_\delta \le Y \}} \Big]\Big\}\\
&\qquad\le Q_{\alpha_1}[(1-\alpha)^{a + 1}-(1-\alpha)^{a + 1}]\\
&\qquad= 0.
\end{align*}

\noindent In the above inequality, we have used
\begin{enumerate}
\item if $Q_\alpha \le S < Q_{\alpha_1}$, then
\begin{align*}
\mathbf{1}_{\{Q_\alpha \le S\le Q_{\alpha_1}\}}-\frac{(1- \alpha)^{a + 1}\mathbf{1}_{\{Q_\alpha \le S \le Q_{\alpha_1},Q_\delta \le Y \le Q_{\delta_1} \}}}{\alpha_1-\alpha-\delta+C(\alpha,\delta)} \le 0;
\end{align*}
\item if $S \ge {Q_{{\alpha _1}}}$, then
\begin{align*}
\mathbf{1}_{\{Q_\alpha \le S\le Q_{\alpha_1}\}}-\frac{(1- \alpha)^{a + 1}\mathbf{1}_{\{Q_\alpha \le S \le Q_{\alpha_1},Q_\delta \le Y \le Q_{\delta_1} \}}}{\alpha_1-\alpha-\delta+C(\alpha,\delta)} \ge 0.
\end{align*}
\end{enumerate}
This proves that MCoVaR has a lower-value than the DCoVaR.
		
\bigskip	
\item We may write the Copula CoVaR as
\begin{align*}
\text{CCoVaR}_\alpha^\delta(S|Y;C)= \frac{1}{1-\alpha-\delta+C(\alpha,\delta)}E[S\, \mathbf{1}_{\{S \ge Q_\alpha,Y \ge Q_\delta\}}].
\end{align*}
Thus,
\begin{align*}
&(1-\alpha-\delta+C(\alpha,\delta))\Big[\text{CCoVaR}_\alpha^\delta(S|Y;C)-\text{DCoVaR}_{(\alpha,a)}^{(\delta,0)}(S|Y;C) \Big]\\
&\qquad=(1-\alpha-\delta+C(\alpha,\delta)) \times \Bigg[\frac{E(S\ \mathbf{1}_{\{S \ge Q_\alpha,Y \ge Q_\delta\}} )}{1-\alpha-\delta+C(\alpha,\delta)}-\frac{E(S\ \mathbf{1}_{\{Q_\alpha \le S \le Q_{\alpha_1},{Q_\delta } \le Y \}})}{\alpha_1-\alpha-\delta+C(\alpha,\delta)} \Bigg]\\
&\qquad \ge Q_{\alpha_1}[(1-\alpha-\delta+C(\alpha,\delta))-(1-\alpha-\delta+C(\alpha,\delta))]\\
&\qquad = 0.
\end{align*}		
In the above inequality, we have used
\begin{enumerate}
\item if $S>Q_{\alpha_1}$, then
\begin{align*}
\mathbf{1}_{\{S\ge Q_\alpha,Y \ge Q_\delta\}}-\frac{(1-\alpha-\delta+C(\alpha,\delta))\mathbf{1}_{\{Q_\alpha\le S \le Q_{\alpha_1},Q_\delta \le Y\}}}{\alpha_1-\alpha-\delta+C(\alpha,\delta)} \ge 0;
\end{align*}
\item if $Q_\alpha\le S \le Q_{\alpha_1},$ then
\begin{align*}
\mathbf{1}_{\{S\ge Q_\alpha,Y \ge Q_\delta\}}-\frac{(1-\alpha-\delta+C(\alpha,\delta))\mathbf{1}_{\{Q_\alpha\le S \le Q_{\alpha_1},Q_\delta \le Y\}}}{\alpha_1-\alpha-\delta+C(\alpha,\delta)} \le 0.
\end{align*}
\end{enumerate}
This proves that DCoVaR has a lower-value than the CCoVaR.	 \end{enumerate}


%
%
%
%



@article{article,
	author = {Jorion, Philippe},
	year = {1996},
	month = {11},
	pages = {47-56},
	title = {Risk 2 : Measuring the Risk in Value at Risk},
	volume = {52},
	journal = {Financial Analysts Journal - FINANC ANAL J},
	doi = {10.2469/faj.v52.n6.2039}
}
\end{document}